\documentclass[prc,twocolumn,showpacs,preprintnumbers,amsmath,amssymb,nofootinbib]{revtex4}

\usepackage{graphicx}
\usepackage{dcolumn}
\usepackage{bm}
\usepackage{hyperref}
\usepackage[mathlines]{lineno}
\usepackage{multirow}

\newcommand{\hfbax}{\sc hfb-ax}
\newcommand{\rr} {\boldsymbol{r}}

\begin{document}

\title{Hartree-Fock-Bogoliubov descriptions of deformed weakly-bound nuclei in large coordinate spaces}
\author{Y.N. Zhang}
\affiliation{State Key Laboratory of Nuclear
Physics and Technology, School of Physics, Peking University,  Beijing 100871, China}
\author{J.C. Pei}
\email{peij@pku.edu.cn}
\affiliation{State Key Laboratory of Nuclear
Physics and Technology, School of Physics, Peking University,  Beijing 100871, China}
\author{F.R. Xu}
\affiliation{State Key Laboratory of Nuclear
Physics and Technology, School of Physics, Peking University,  Beijing 100871, China}

\begin{abstract}
Weakly-bound deformed nuclei have been studied by the Skyrme Hartree-Fock-Bogoliubov (HFB) approach in large coordinate-space boxes.
In particular, the box-size dependence of the HFB calculations of weakly-bound deformed nuclei are investigated,
including the particle density and pairing density distributions at nuclear surfaces, the near-threshold resonant and continuum quasiparticle
spectra, and energetic properties. The box size may have larger influences in pairing
 properties than in other bulk properties. We demonstrate that large-box calculations of weakly-bound nuclei
 are important to precisely describe exotic phenomena such as deformed halos and peninsulas of stability beyond drip lines.

\end{abstract}

\maketitle


\section{Introduction}

The nuclei close to the particle drip lines are very weakly bound and can have exotic properties according
to our knowledge of stable nuclei, which makes them fascinating quantum systems exhibiting the threshold effects, such
as halo states~\cite{halo-exp,halo-87,halo,halo-rev}. There have been numerous theoretical studies on the halos of weakly bound nuclei, but most of them are spherical~\cite{Mizutori,schunck,rotival}.
In particular, for weakly-bound deformed nuclei,  exotic deformed halos with decoupled surface deformations from the cores are predicted~\cite{misu,sgzhou,pei2013}. It was known that the continuum degree of freedom plays an important role in weakly-bound nuclei~\cite{continuum-jacek,continuum,forssen,hagen,pei2013}. Therefore, the theoretical descriptions of weakly-bound deformed nuclei should precisely
take into account the continuum effect, deformations and large spatial extensions. In this context, the self-consistent HFB approach of density functional theory with continuum couplings is a suitable method. However, the exact
treatment of continuum states in the HFB approach for deformed nuclei is rare and much more complicated
compared to the spherical case~\cite{matsuo2009}.

In the HFB approach, the continuum can be treated either by the discretization method or by exact solutions with scattering boundary conditions ~\cite{Michel,matsuo2009}.
In the discretization method, the continuum can be discretized on a discrete set of basis functions~\cite{hfbtho,hfodd} or on the coordinate-space lattice~\cite{hfbrad,teran}.
Conventionally, the HFB solvers are based on the HO basis which are very efficient but not suitable for describing weakly-bound systems.
 It has been demonstrated that the coordinate-space HFB approach
is very precise for describing weakly-bound nuclei and continuum effects~\cite{hfbrad,teran,pei2011}.
For halo structures with large spatial extensions, the coordinate-space HFB descriptions definitely need
large box sizes. In addition, it was known that the resulted number of continuum states increases significantly ($\propto L^3$)
as the box size $L$ increases~\cite{continuum-jacek}. The very dense quasiparticle energy spectrum can provide good resolutions for the resonance and
non-resonant continuum. Therefore it will be very useful to study the weakly-bound deformed nuclei by the HFB approach in large boxes.
However, the computational cost will
be increased tremendously as the box sizes is increased in deformed cases.  Fortunately, due to the development of supercomputing capabilities,
the deformed HFB descriptions in large box can be realized through
large-scale parallel calculations~\cite{pei2012}.

In deformed weakly-bound nuclei, the subtle interplay
among surface deformations, surface diffuseness, and continuum coupling can result in exotic structures, and the theoretical studies need
precise HFB solutions. For example, there was an argument about the existence of two-neutron deformed halos based on the three-body model~\cite{nues}.
while the existence of one-neutron deformed halos has been confirmed experimentally~\cite{Ne31,halo-rev},
In our previous work~\cite{pei2013}, we have shown that large-box HFB calculations give
 an exotic ``egg"-like halo structure in $^{38}$Ne showing the core-halo deformation decoupling, which is associated with the phase space decoupling in the quasiparticle energy spectra. It was demonstrated that the near-threshold non-resonant continuum is mainly responsible for such an exotic halo structure~\cite{pei2013}. Thus the quasiparticle spectrum provides an important test for the precision of HFB solutions.
 Some earlier studies showed that calculations with small box sizes may not be sufficient to describe the pairing properties~\cite{grasso2001} and quasiparticle spectrum of pairing occupation numbers~\cite{matsuo2009}. Besides, one may be cautious about the continuum discretization method for descriptions of broad quasiparticle resonances as the reaction model CDCC has been suffering~\cite{CDCC}.  It is also important to evaluate the influences of the precision of HFB solutions in the prediction of drip lines.
  Indeed, the usually adopted accuracy benchmark of HFB calculations by comparing total binding energies may not be sufficient for descriptions of weakly-bound nuclei in details.
   Our motivation in this paper
is to further examine the HFB calculations of weakly bound deformed nuclei and study the box size dependence in several aspects,
including energetic properties, deformations, densities and pairing densities, and the near-threshold quasiparticle spectra of resonances and continuum,
 with much larger box sizes compared to our previous work. This
will be useful for further studies of novel structures and excitation modes in weakly-bound deformed nuclei so that one is sure that they do not arise from
uncontrolled approximations.

\section{Theoretical Method}

In this work, the Skyrme-HFB equation is solved by the {\hfbax} code~\cite{Pei08,pei2011,pei2012} within a 2D lattice box, based on B-spline techniques for
axially symmetric deformed nuclei~\cite{teran}. To obtain sufficient accuracy, the adopted 2D box size is very large up to 36$\times$36 fm.
 The maxima mesh size is 0.6 fm and the order of B-splines is 12.
This is the first deformed HFB calculations with such a large box size,
 while in our previous work the adopted 2D box size is 30$\times$30 fm~\cite{pei2013}.
For calculations employing large boxes and small lattice spacings,
 the discretized continuum spectra would be very dense and provide good resolutions.
Because the computing cost is extremely high, the hybrid MPI+OpenMP parallel programming is implemented to get
converged results within a reasonable time~\cite{pei2012}. It has to be noted that the {\hfbax} code has
been improved remarkably for descriptions of large systems~\cite{pei2012}, compared to its initial version~\cite{Pei08}.
Calculations were performed in China's top supercomputer Tianhe-1A.
For the particle-hole interaction channel, the SLy4 force~\cite{sly4} is adopted as
it is one of the mostly used parameterizations for neutron rich nuclei.
For the particle-particle channel, the density dependent surface pairing interaction
is used~\cite{mix-pairing}. The pairing strengthes are fitted to the neutron gap of $^{120}$Sn.

The HFB equation in the coordinate-space representation can be written as

\begin{equation}\label{1}
\left[
  \begin{array}{cc}
    h(\rr)-\lambda & \Delta(\rr) \\
    \Delta^{*}(\rr) & -h(\rr)+\lambda \\
  \end{array}
\right]
  \left[
    \begin{array}{c}
      U_{k}(\rr) \\
      V_{k}(\rr) \\
    \end{array}
  \right]
  = E_k
    \left[
    \begin{array}{c}
      U_{k}(\rr) \\
      V_{k}(\rr) \\
    \end{array}
  \right],
\end{equation}
where $h$ is the Hartree-Fock Hamiltonian; $\Delta$ is the pairing potential;
$U_k$ and $V_k$ are the upper and lower components of quasi-particle
wave functions, respectively; $E_k$ is the quasi-particle energy; and
$\lambda$ is the Fermi energy (or chemical potential).
For bound systems, $\lambda < 0$ and the self-consistent densities and fields
are localized in space.
For $|E_k|<-\lambda$, the eigenstates are discrete and
$V_k(\rr)$ and $U_k(\rr)$ decay exponentially.
The quasiparticle continuum corresponds to  $|E_k|>-\lambda$. For those states, the upper component of the
wave function always has  a scattering asymptotic form. By applying the
box boundary condition, the continuum becomes discretized and one obtains
a finite number of continuum quasi-particles. In principle, the box solution
representing the continuum can be  close to the exact
solution when a sufficiently big box and small mesh size are adopted.

Based on the quasiparticle wave functions, the particle density $\rho(\mathbf{r})$ and the pairing density $\tilde{\rho}(\mathbf{r})$ can be written as
\begin{equation}
\begin{array}{c}
\rho(\mathbf{r})=\sum_k V_k^{*}(\rr)V_k(\rr) \vspace{3pt}\\
\tilde{\rho}(\mathbf{r})=-\sum_k V_k(\rr)U_k^{*}(\rr)
\end{array},
\end{equation}
where in the sum the quasiparticle energy cutoff is taken as ($60-\lambda$) MeV. We also discussed
the particle occupation numbers $n_k$ and pairing occupation numbers $\tilde{n}_k$  as defined below:
\begin{equation}
\begin{array}{c}
n_k = \int V_k^{*}(\rr)V_k(\rr) d^3 \rr  \vspace{3pt}\\
\tilde{n}_k=-\int V_k(\rr)U_k^{*}(\rr)  d^3 \rr
\end{array}.
\label{occupationn}
\end{equation}

\section{Calculations  and discussions}

\subsection{Density and pairing density distributions at nuclear surfaces}

Firstly we study the particle density and pairing density distributions of weakly-bound deformed nuclei.
Here we focus on the deformed Ne and Mg isotopes near the neutron drip line.
Experimentally, $^{34}$Ne and $^{40}$Mg are the neutron-richest isotopes known experimentally so far~\cite{nature-mg}.
Theoretically, there is a possibility of deformed neutron halos in this region~\cite{sgzhou,pei2013}.

\begin{figure}[t]
\centerline{\includegraphics[trim=0cm 0cm 0cm
0cm,width=0.5\textwidth,clip]{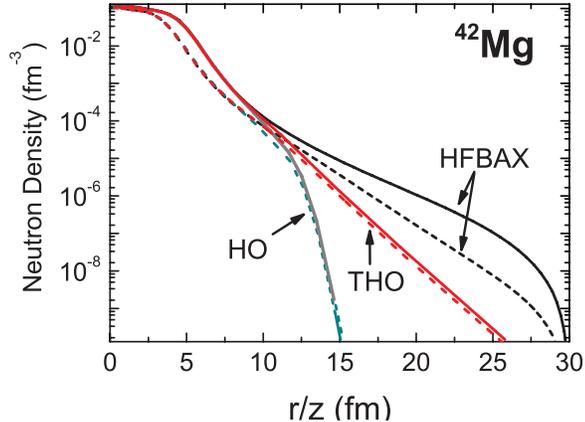}}
  \caption{(Color online) The neutron density profiles of $^{42}$Mg  are calculated by HFB-HO, HFB-THO and HFB-AX. The HFB-HO and HFB-THO calculations are based on 30 HO shells and the HFB-AX calculations are done within a box of 30 fm. The density distributions are displayed along the cylindrical coordinates $z$-axis(solid line) and $r$-axis(dashed line), respectively. }
  \label{mg42}
\end{figure}

In Fig.\ref{mg42} we display the neutron density profiles of $^{42}$Mg obtained by different solving methods.
The comparison between the HFB solutions based on the Harmonic Oscillator (HO) basis, the transformed HO (THO) basis~\cite{hfbtho} and the coordinate-space HFB solutions of {\hfbax} are shown in logarithmic scale to emphasize different surface asymptotics. The densities are
displayed along
the cylindrical  coordinates $z$-axis (the axis of symmetry)  and $r$-axis (the axis perpendicular to $z$-axis and $r$=$\sqrt{x^2+y^2}$), respectively.
The differences between the density profiles $\rho_{z({r=0})}$ and  $\rho_{r({z=0})}$  actually reflect the surface deformations.
It can be seen that the results obtained from three methods have very different asymptotic behaviors at large distances. With the HO basis, the density distributions decays very rapidly
due to the Gaussian asymptotic of HO basis.
For this, the THO basis aims to improve the descriptions of weakly-bound nuclei compared to HO basis~\cite{mario03}. However, THO basis calculations still fail to
 reproduce the halo structure and the surface deformation in $^{42}$Mg compared to the coordinate-space calculations.
Based on the comparison, the accuracy and advantages of coordinate-space calculations have been clearly illustrated for descriptions of weakly-bound deformed nuclei.

\begin{figure}[t]
  \includegraphics[width=0.45\textwidth]{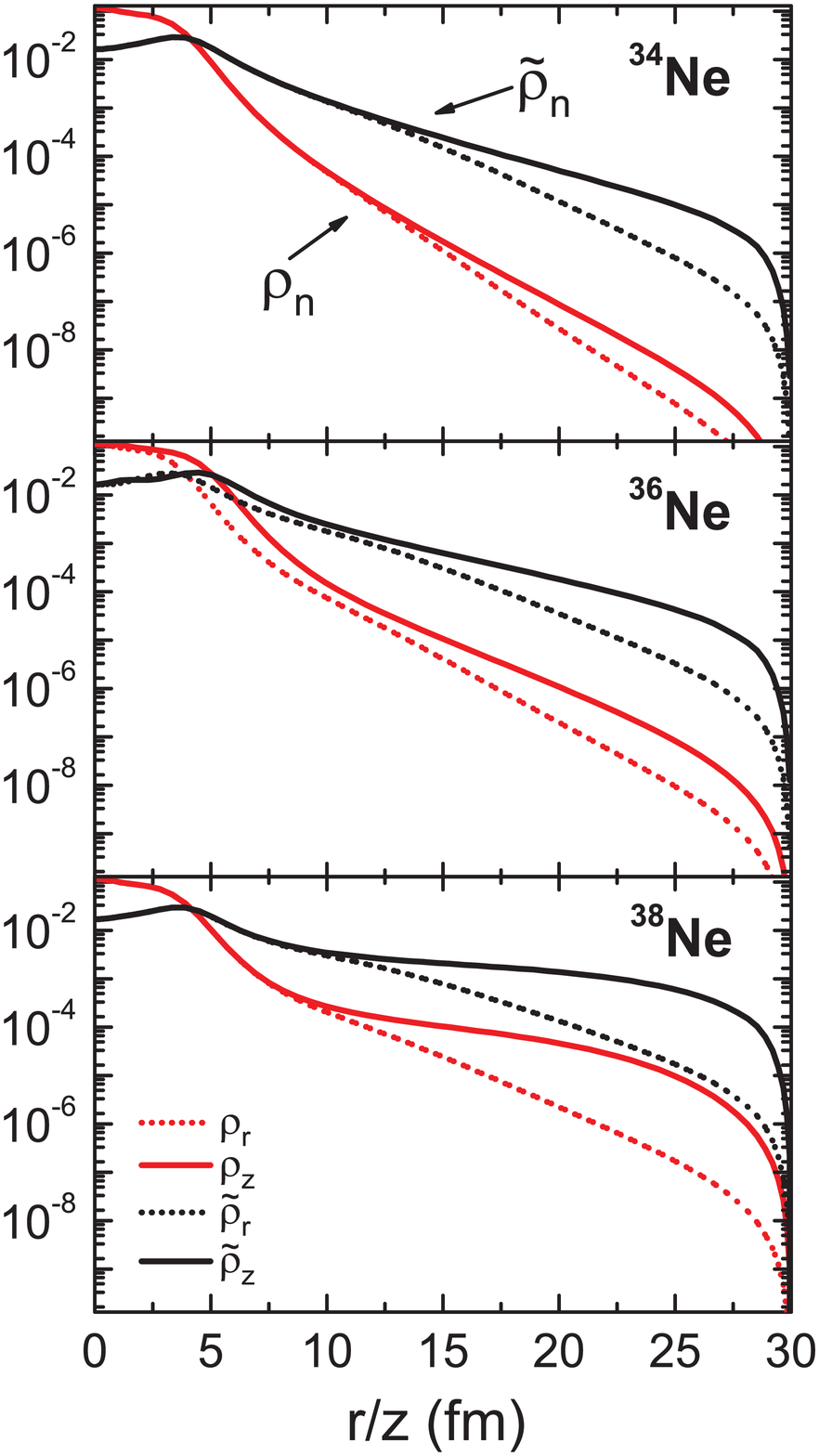}\\
  \caption{(Color online) The neutron density and neutron pairing density profiles of $^{34}$Ne,$^{36}$Ne and $^{38}$Ne. The density $\rho$ and pairing density $\tilde{\rho}$ distributions are displayed along the cylindrical coordinates $z$-axis (solid line) and $r$-axis (dashed line), respectively. }
  \label{neiso}
\end{figure}

\begin{figure}[t]
  \includegraphics[width=0.45\textwidth]{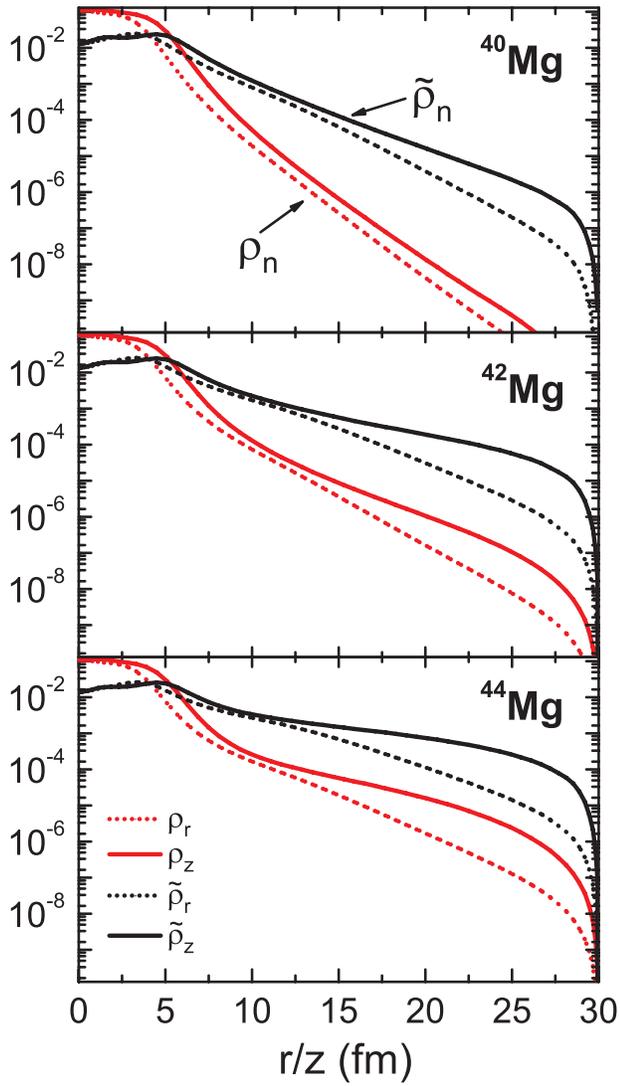}\\
  \caption{(Color online) The neutron density and neutron pairing density profiles of $^{40}$Mg,$^{42}$Mg and $^{44}$Mg. The density $\rho$ and pairing density $\tilde{\rho}$ distributions are displayed along the cylindrical coordinates $z$-axis(solid line) and $r$-axis(dashed line), respectively. }
  \label{mgiso}
\end{figure}

In Figs. \ref{neiso} and \ref{mgiso} we show the neutron density distributions of Ne and Mg isotopes near the neutron drip line.
It can be seen that as the neutron number increase, in general, the surface densities enhance and the halo structures become more pronounced.
In Ne isotopes, the surface deformations have interesting evolutions. In Fig.\ref{neiso},  $^{34}$Ne has a spherical core and a small surface deformation;
$^{36}$Ne has a deformed core and a small surface deformation; $^{38}$Ne has a spherical core plus a well deformed prolate halo.
Such an ``egg''-like halo structure in $^{38}$Ne has been pointed out in the previous work as a result of
the subtle interplay between the surface diffuseness, surface deformation and continuum couplings~\cite{pei2013}.
In Mg isotopes, all the cores and the skins/halos are well prolate deformed. It has to be noted that the surface deformations increase in both Ne and
Mg isotopes close to the neutron drip line.  It was known that the surface deformation and the surface diffuseness are mainly contributed by
the near-threshold non-resonance continuum~\cite{pei2013}. Therefore the increase of surface deformations can be attributed to the enhanced non-resonant
continuum close to the neutron drip line.
In Figs. \ref{neiso} and \ref{mgiso}, we have also displayed the neutron pairing density distributions $\tilde{\rho}_r$ and $\tilde{\rho}_z$ of Ne and Mg isotopes.
As we have pointed out earlier, the deformation decoupling also occurs in the pairing density distributions~\cite{pei2013}.
In the spherical case, the much larger spatial extensions of the pairing densities have
been observed in drip line nuclei, due to the different asymptotic behaviors of $\rho$ and $\tilde{\rho}$ at large distances~\cite{continuum-jacek}.
In the deformed cases, it can be seen that not only the spatial extensions but also the surface deformations of pairing density distributions are
larger than that of the density distributions.  The deformed halo structures are more significant in the pairing density distributions.
This can also be interpreted as that the non-resonant continuum has much larger influences in pairing properties than
in normal densities~\cite{pei2011,zhang}.

\begin{figure}[htb]
  \includegraphics[width=0.45\textwidth]{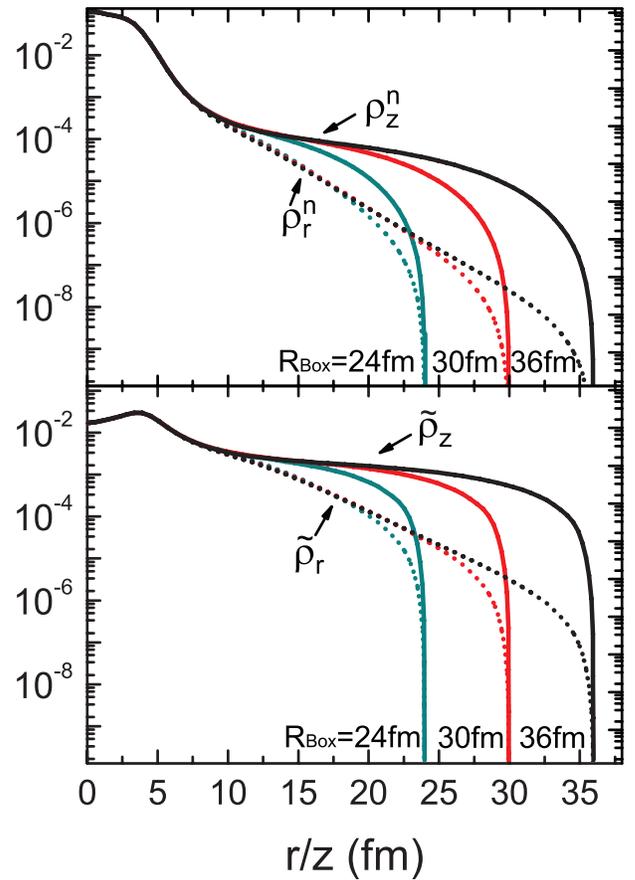}\\
  \caption{(Color online) The neutron density $\rho_{n}$ (top), and neutron pairing density $\tilde{\rho}_{n}$ (bottom) of $^{38}$Ne calculated with box sizes of 24 fm, 30 fm and 36 fm. The density and pairing density distributions are displayed along the cylindrical coordinates $z$-axis (solid line) and $r$-axis (dashed line), respectively. }
  \label{denbox}
\end{figure}

Fig. \ref{denbox} displays the neutron density $\rho(r)$ and the neutron pairing density $\tilde{\rho}(r)$ of $^{38}$Ne calculated with the box sizes of 24 fm, 30 fm and 36 fm, respectively. It can be seen that, for different box sizes, the densities have the same asymptotics and surface deformations before meeting the box boundaries.  This implies the
deformed halo structure of $^{38}$Ne is rather robust as the box size changes.
We can see that the box size of 30 fm is sufficient to give rise to the deformed halo in the density distributions.  For the pairing density distribution with
a much larger spatial extension, however, we can see that the HFB calculations require an even larger box.

\subsection{Quasiparticle spectrum near thresholds}

It was known that the near-threshold quasiparticle spectrum, in particular the
non-resonant continuum,  is mainly responsible for the halo structures in weakly-bound nuclei~\cite{forssen,zhang,pei2013}.
While the conventional interpretation of halos in terms of weakly-bound single-particle wavefunctions could be oversimplified
for the two-neutron halos.
Therefore it is of great interest to study the structures of quasiparticle spectrum of resonances and continuum, especially for
deformed weakly-bound nuclei.

\begin{figure}[t]
  \includegraphics[width=0.45\textwidth]{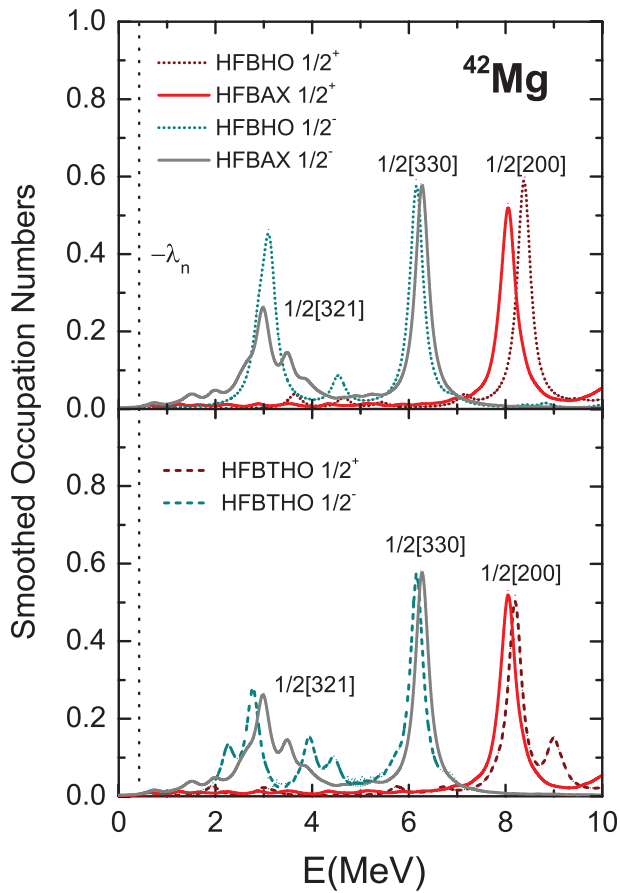}\\
  \caption{(Color online) The smoothed particle occupation numbers of $\Omega^{\pi}=1/2^{\pm}$ neutron quasiparticle states of $^{42}$Mg,
  The quasiparticle spectra obtained by  HFB-HO (dotted line) and HFB-THO (dashed line) with 30 shells of HO/THO basis, and by HFB-AX (solid line) with a box size of 30 fm. }
  \label{smooth1}
\end{figure}

To compare the different quasiparticle spectra calculated by the HO basis, THO basis and the coordinate-space discretization method,
we display the smoothed occupation numbers $\tilde{n}_i$ [as defined in Eq.(\ref{occupationn})], as showed in Fig.\ref{smooth1}.
The neutron quasiparticle occupation numbers $n_{i}$ of $\Omega^{\pi}=1/2^{\pm}$ of $^{42}$Mg are smoothed with a Lorentz shape function and a smoothing parameter of $50 keV$.
The smoothing method was described in Ref.\cite{pei2011}.
For quasiparticle resonances, the related Nillsson labels are given.
The quasiparticle spectra from HFB-HO have a significant shift compared to HFB-AX, and the THO basis shows improved agreement compared to HFB-AX.
Indeed, HFB-THO has better surface asymptotics than HFB-HO for $^{42}$Mg as shown in Fig.\ref{mg42}.
However, both the HO and THO calculations have problems to represent the 1/2[321] which is a broad quasiparticle resonance.
The HFB-HO calculations tends to underestimate the widths of broad resonances.
It is known that the discretized spectra of quasiparticle resonances can roughly have the Breit-Wigner shape~\cite{pei2011}.
For quasiparticle states with narrow widths, the basis expansion
method should be fine. It is obvious that the coordinate-space method is superior over the HO or THO basis expansion method
in describing the broad quasiparticle resonances. This is because the coordinate-space HFB calculations produce much denser quasiparticle spectra than
the HO and THO calculations.

\begin{figure}[t]
  \includegraphics[width=0.45\textwidth]{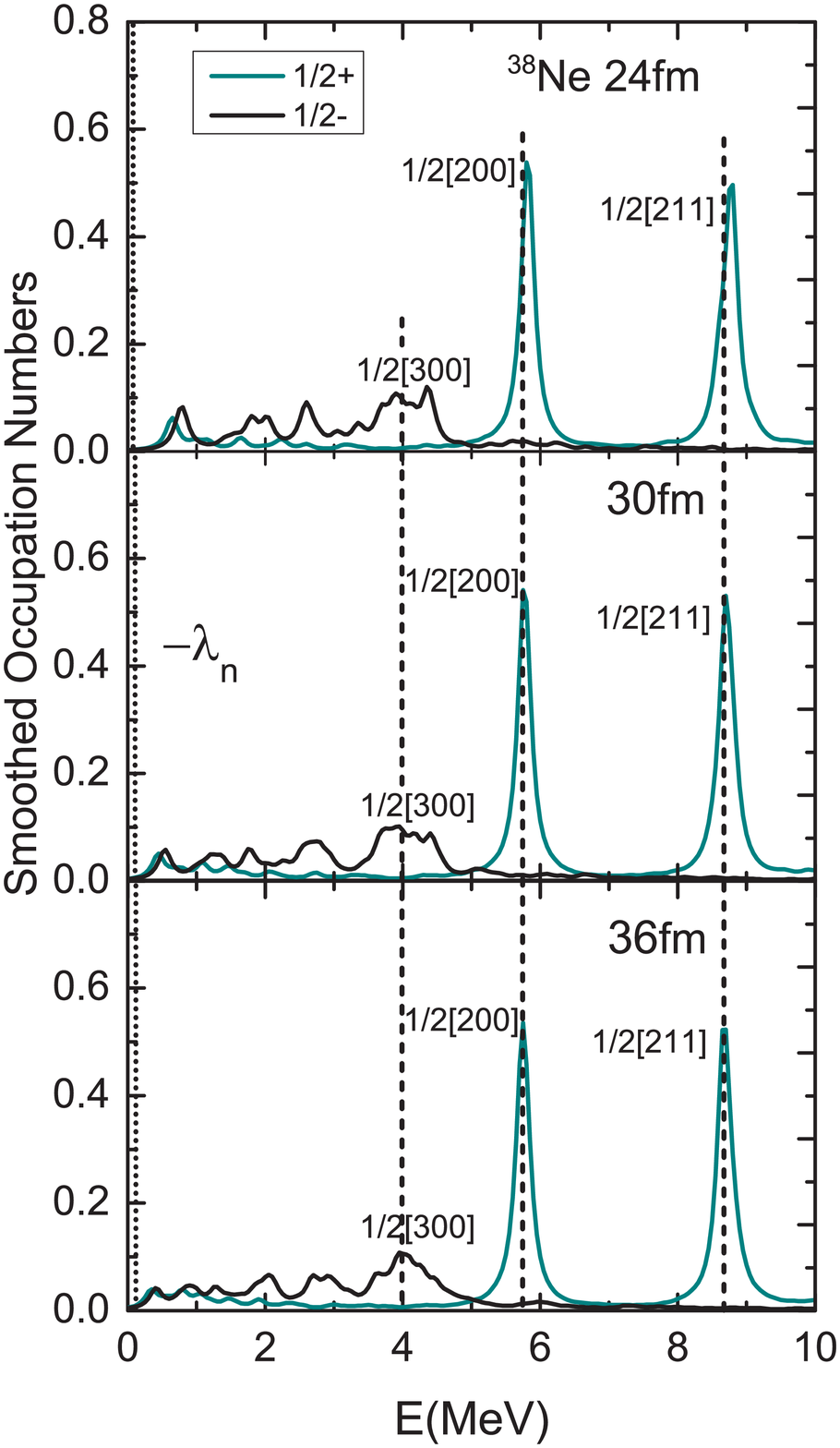}\\
  \caption{(Color online) The smoothed particle occupation numbers of $\Omega^{\pi}=1/2^{\pm}$ neutron quasiparticle states of $^{38}$Ne obtained by HFB-AX with box sizes of 24 fm, 30 fm, and 36 fm. The continuum thresholds, i.e. $-\lambda_{n}$, are given, where $\lambda_{n}$ is the neutron Fermi energy.}
  \label{ne38qp1}
\end{figure}

Next we demonstrate that larger box calculations can further improve the descriptions of broad quasiparticle resonances and non-resonant continuum.
To distinguish the quasiparticle resonances and the continuum, we display the smoothed neutron quasiparticle spectra of $^{38}$Ne obtained with box sizes of 24fm, 30fm and 36fm,
as shown in Fig.\ref{ne38qp1}.
It was known that the energies of quasiparticle resonances are stationary with different box calculations, while continuum states are not~\cite{doba1984}.
With box-changed calculations one can get the information about non-resonant continuum contributions.
It is seen that the quasiparticle spectrum above the thresholds consisting of several resonances,
and a remarkable non-resonant continuum background below 3 MeV. In our previous study~\cite{pei2013}, the near-threshold non-resonant continuum is mainly responsible
for the ``egg''-like deformed halo structure in $^{38}$Ne.
In Fig.\ref{ne38qp1}, the distributions of 1/2[200] and 1/2[211] states don't change as the box size varies and they are obviously quasiparticle  resonances.
However, it is difficult to identify the broad resonance 1/2[300] around 4 MeV. By increasing the box sizes, it can be seen that
the resonance 1/2[300] are stationary and becomes more and more like a Breit-Wigner shape. There are some small peaks close to the threshold that fades away
as the box size increases. Actually they should be completely dissolved into the continuum in even larger box calculations
or exact HFB solutions.

\begin{table}[htb]
\caption{\label{ne38tab}
 Calculated results of $^{38}$Ne obtained by HFB-AX with box sizes of 24 fm, 27 fm, 30 fm and 36fm.
 The total binding energy $E_{tot}$, Coulomb energy $E_c$, pairing energy $E_{pair}$, kinetic energies $E_{kin}$,
 the quadruple deformations $\beta_2$, the r.m.s radii $R_{rms}$ (in fm), the pairing gaps $\Delta$, the Fermi surface energies $\lambda$
 are listed. The energies are given in MeV.}
\begin{tabular}{lrrrrrrr}
\hline
\hline\\[-0.7em]
\vspace{0.2mm}
~~~~$^{38}$Ne         &~ 24fm &~ 27fm &~ 30fm &~ 36fm  \\
\hline\\[-0.7em]
~~ $E_{tot}$     &~~ $-$220.29~   &~~$-$220.29   &~~ $-$220.35   &~~ $-$220.33  \\
~~  $E_{c}$      &~~18.94~      &~~18.95      &~~18.95      &~~18.96      \\
~~  $E_{pair}$ &~~ $-$68.10~    &~~ $-$67.77    &~~ $-$67.41    &~~ $-$67.43    \\
~~  $E^{p}_{kin}$       &~~ 138.34~    &~~ 138.46    &~~ 138.51    &~~ 138.56    \\
~~  $E^{n}_{kin}$       &~~ 444.84~    & ~~443.48    &~~ 442.85    &~~442.10    \\
~~  $\beta_{2p}$  &~ ~0.00~      &~~ 0.00      &~~ 0.00      &~~ 0.00      \\
~~  $\beta_{2n}$  &~~ 0.13~      &~~ 0.19      &~~ 0.24      &~~ 0.34      \\
~~  $R_{rms}$     &~ 9.39~      &~ 9.38      &~ 9.38      &~ 9.37      \\
~~  $\Delta_{p}$  &~~ 1.46~      & ~~1.47      &~ ~1.47      &~ ~1.46     \\
~~  $\Delta_{n}$  &~~ 2.97~      &~ ~2.95      &~~ 2.93      &~ ~2.92      \\
~~  $\lambda_{p}$ & $-$23.853   &~ $-$23.802   &~ $-$23.784    &~ $-$23.747    \\
~~  $\lambda_{n}$ & $-$0.079    &~ $-$0.096    &~ $-$0.103    &~ $-$0.116    \\
  \hline
  \hline
\end{tabular}
\end{table}

\begin{figure}[htb]
  \includegraphics[width=0.45\textwidth]{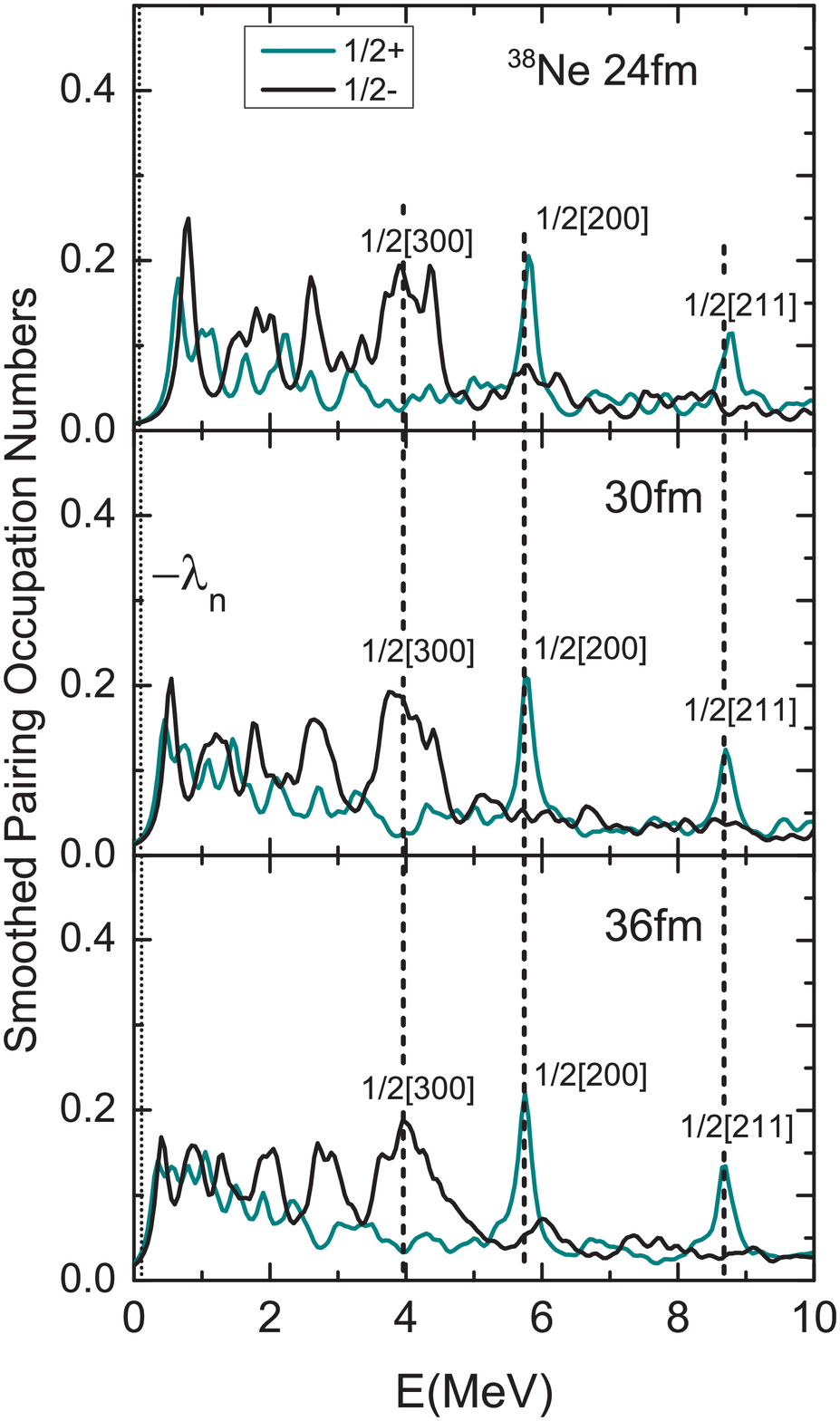}\\
  \caption{(Color online) The smoothed pairing occupation numbers of $\Omega^{\pi}=1/2^{\pm}$ neutron quasiparticle states of $^{38}$Ne obtained by HFB-AX with box sizes of 24 fm, 30 fm, and 36 fm.
  The continuum thresholds, $-\lambda_{n}$, are given, where $\lambda_{n}$ is the neutron Fermi energy.}
  \label{ne38qp2}
\end{figure}

Fig. \ref{ne38qp2} displays the smoothed pairing occupation numbers $\tilde{n}_i$ obtained by different box sizes, corresponding to the particle occupation numbers $n_i$ in Fig.\ref{ne38qp1}. We can see that the near-threshold non-resonant continuum  have significant contributions that exceed the contributions of resonances.
The dominance of the non-resonant continuum in the pairing channel of weakly-bound nuclei has been pointed out in \cite{zhang,forssen}.
For each quasiparticle resonance, the spectrum of pairing occupation numbers has a corresponding peak.
For the broad resonance 1/2[300], it can be seen that the description of the pairing occupation numbers can be improved by increasing the box size.
The failure of descriptions of pairing occupation numbers in small box solutions has been pointed out by~\cite{matsuo2009}.
In Fig. \ref{ne38qp2}, the non-resonant part of the pairing occupation numbers still has notable non-smooth distributions even with a box size of 36 fm.
Based on the comparison between Fig. \ref{ne38qp1} and Fig. \ref{ne38qp2}, one can understand that even larger box sizes are needed for descriptions of pairing properties, especially
for the surface peaked pairing shown in Fig.\ref{denbox}.

Table \ref{ne38tab} shows the result of deformed HFB calculations for $^{38}$Ne with different box sizes.
With the increasing box size, the total binding energy $E_{tot}$ increases generally , indicating the stability is being enhanced.
The pairing energy $E_{pair}$ and the neutron kinetic energy $E_{kin}$ decrease significantly as the box size increases. The combination $E_{pair}+E_{kin}$
is less sensitive to the box sizes.
It was pointed out in \cite{grasso2001} that in the vicinity of the drip lines pairing correlations can be overestimated by the continuum discretization in box HFB calculations.
Indeed, the convergence of the pairing energy is slow with respect to the increasing box sizes. Again this indicates that large box calculations
are essential for describing pairing properties of weakly-bound nuclei and exotic deformed halo structures. We also observed
the increase in the surface quadrupole deformations $\beta_{2n}$ of neutrons as the box size increases, although they have consistent surface asympotics as shown in Fig.\ref{denbox}. Nevertheless the deformations in this region are very soft~\cite{terasaki}.
The decreasing of the neutron Fermi surface energy with increasing box sizes implies the halo structure becomes more stable in larger box calculations,
mainly due to the better treatment of the continuum effects and pairing correlations.

\subsection{Peninsula of stability beyond the drip line}

Once the accurate HFB descriptions of weakly-bound deformed nuclei are realized, it is interesting to
address the question whether there can possibly exist islands or peninsulas of stability beyond the neutron drip line~\cite{nazarewicz99}.
There have been some studies based on Hartree-Fock+BCS method to explore the islands or peninsulas of stability~\cite{tarasov}.
Also we like to know the influences of the HFB solving precision on the prediction of the neutron drip line.
We are particularly interested in the possible islands or peninsulas of stability due to deformations and continuum effects
based on the coordinate-space HFB calculations.
Note that the HFB-HO and HFB-THO calculations even with 30 shells of basis still can not obtain the deformed halos in $^{38}$Ne and $^{44}$Mg.
To examine the HFB precision in the heavy mass region, we studied the well deformed Nd ($Z$=60) isotopes with the SLy4 force and
the mixed pairing interaction~\cite{mix-pairing}.

\begin{figure}
  \includegraphics[width=0.45\textwidth]{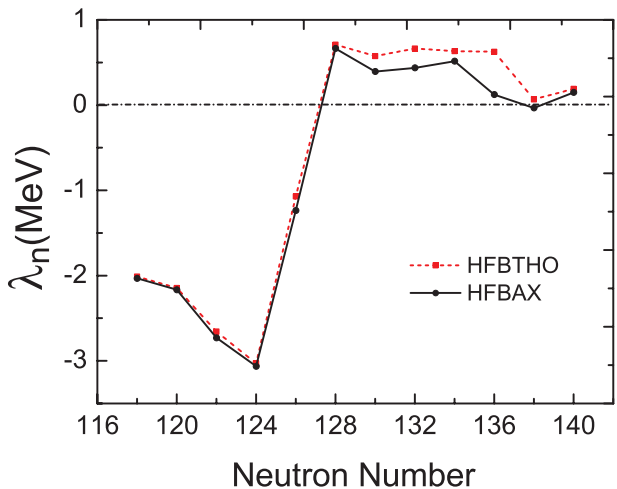}\\
  \caption{(Color online) The neutron Fermi energies of Nd isotopes calculated by HFB-THO (with 30 shells of transformed HO basis) and HFB-AX (with a 30 fm box).  }
  \label{nd-iso}
\end{figure}

In Fig.\ref{nd-iso}, the Fermi surface energies of Nd isotopes are shown as a function of neutron numbers.
It can be seen that the normal two-neutron drip line is obtained by different methods as $N$=126.
The Nd isotopes from $N$=128 to 136 with positive Fermi surface energies are not bound systems.
It is reasonable that the Fermi surface energies in HFB-AX calculations are systematically lower than that in
HFB-THO calculations.
The interesting point is that the Nd isotope with $N$=138 is slightly bound in the coordinate-space HFB calculations, with
a negative $\lambda_n$ of $-$35 KeV. While $\lambda_n$ is $68$ KeV in the HFB-THO calculations with 30 shells of transformed HO basis.
By using the approximation $S_{2n}(Z,N)\approx -2\lambda_{n}$ (for even $N$)~\cite{erler}, $^{198}$Nd seems to be bound for two-neutron emissions.
Hence $^{198}$Nd may be seen as the extension of the peninsula of stability in the deformed Nd-Sm-Gd region around $N$=138~\cite{erler}.
The nucleus $^{198}$Nd is well deformed ($\beta_2$=0.28) and has considerable pairing correlations ($\Delta_n$=0.59 MeV) as well as continuum effects.  It has to be noted that, however, its halo structure
is not so significant as that in light nuclei. Calculations with mixed pairing indeed give arise to less significant halo features compared to the surface pairing.
In addition, the decreased possibility of heavy halos has been discussed in Refs.\cite{pei2013,schunck,rotival}.
With the Nd example studied, we might predict the possibilities of other islands or peninsulas of stability by accurate HFB calculations.

\section{Summary}

The main objective of this study is to perform detailed analysis of  deformed weakly bound nuclei based on the self-consistent Skyrme-HFB approach within large coordinate-space boxes. The advantages of the coordinate-space HFB calculations are its capability of the precisely treatment
of the surface deformations, continuum effects and large spatial extensions, which are
essential mechanisms of the exotic structures in weakly-bound deformed nuclei.
Based on detailed analysis and the comparison with the HO and THO basis expansion methods, we demonstrated that large box HFB calculations
are necessary to describe the asymptotic behaviors of deformed halos and pairing density distributions,  broad quasiparticle resonances and non-resonant continuum,
the predictions on islands or peninsulas of stability and drip lines. In addition, the large box calculations are especially needed for
descriptions of pairing occupation numbers and pairing density distributions of weakly-bound nuclei which have more pronounced deformed halo structures.
This is related to the fact that the pairing density distributions have lager surface deformations and larger spatial extensions than that of the
normal densities. For this, the remarkable contribution from near-threshold non-resonant continuum plays an important role.
One should keep in mind that the precise predictions of deformed halos would be dependent upon
models and effective interactions that one used.

\section*{Acknowledgments}
 This work was supported by the
National Key Basic Research Program of China under Grant 2013CB834400,
and the National Natural Science Foundation of China under Grants No.11375016, 11235001 and 11320101004.
We also acknowledge that computations in this work were performed in the Tianhe-1A supercomputer
 located in the Chinese National Supercomputer Center in Tianjin.

\nocite{*}


\end{document}